\documentclass[aps,prb,superscriptaddress,twocolumn]{revtex4-2}
\usepackage[margin=1in]{geometry}
\usepackage{graphicx} 
\usepackage{amsmath,amssymb,bm}
\usepackage{url}
\usepackage[colorlinks=true]{hyperref}
\usepackage{caption}
\usepackage{setspace}
\usepackage{titlesec}
\usepackage{indentfirst}
\usepackage{xcolor}
\usepackage{siunitx}
\usepackage{xr}
\begin{document}

\title{Tunable anisotropy of high-harmonic generation in ferroelectric NbOI$_2$}

\author{Inho Lee}
\affiliation{Theoretical Division, Los Alamos National Laboratory, Los Alamos, New Mexico 87545, USA}

\author{Christopher Lane}
\affiliation{Theoretical Division, Los Alamos National Laboratory, Los Alamos, New Mexico 87545, USA}

\author{Daniel A. Rehn}
\affiliation{Computational Physics Division, Los Alamos National Laboratory, Los Alamos, New Mexico 87545, USA}

\author{Jian-Xin Zhu}
\email{jxzhu@lanl.gov}
\thanks{Corresponding author}
\affiliation{Theoretical Division, Los Alamos National Laboratory, Los Alamos, New Mexico 87545, USA}
\affiliation{Center for Integrated Nanotechnologies, Los Alamos National Laboratory, Los Alamos, New Mexico 87545, USA}

\date{\today}

\begin{abstract}
The generation of even-order harmonic spectra from inversion symmetry breaking is well known, however, the tunability of the intensity and angular anisotropy on various harmonics orders remains unclear. Here, we demonstrate the second and fourth-harmonics in the ferroelectric NbOI$_2$ exhibit distinct wavelength-dependence and anisotropies in the near-infrared regime. In particular, as the wavelength of incident light is tuned from 2150 nm to 2450 nm,
the second-harmonic remains strongest for excitation along the ferroelectric polar axis, where as the preferred axis for fourth-harmonic generation changes with wavelength. This effect is found to be driven by the interplay of transition dipole matrix elements and interband excitations between very narrow bands near the Fermi level.  
We expect the tunable anisotropy of high-harmonics to enable novel optical switching devices.
\end{abstract}

\maketitle

\section{Introduction}

Nonlinear optics in condensed matter systems connects strong-field optical responses to the symmetry, electronic structure, and dynamical properties of matter\cite{Aslam2023,Jauregui2016}.
In solids, high-harmonic generation (HHG) is widely used to examine interplay of electron dynamics in the electronic band structures, Berry curvature, and crystal symmetry. Moreover, HHG can be applied as extreme-ultraviolet source with high conversion efficiency\cite{Ghimire2019} to characterize defects\cite{Cavaletto2025} and probe optically driven phase transition in correlated materials\cite{Silva2018}.

The crystal symmetry, in particular, plays a crucial role in determining the high-harmonic generation spectra, as evidenced by the absence of even-order harmonics in centrosymmetric systems. In crystals that preserve inversion symmetry, even-order harmonics cancel out, thereby restricting the spectrum to strictly odd harmonics. Even harmonics can be recovered by breaking the inversion symmetry via strain\cite{Schriever2010}, surface nonlinearities\cite{Timbrell2018}, and various nano-structures. However, in spite of these symmetry breaking approach, the resulting even-order harmonics are very weak, with experimental observations limited to the lowest orders\cite{Tonkaev2024}.

Non-centrosymmetric materials naturally exhibit sizable even-order harmonics due to the intrinsic lack of inversion symmetry\cite{Ghimire2019,Yue2022}. Second harmonic generation (SHG) and higher even-order harmonics have been observed in several non-centrosymmetric materials including monolayer MoS$_2$ \cite{Saynatjoki2017}, ZnO \cite{Jiang2019}, and GaAs \cite{Stanton2020}. Of these, two material classes have been of particular importance: two-dimensional (2D) van der Waals (vdW) and polar material systems. 2D van der Waals materials, including transition metal dichalcogenides and hexagonal boron nitride have attracted significant attention due to their unique optical properties including valley-selective excitations, excitonic effects, and polaritons with moire-induced nonlinearity\cite{Zhang2021,Kim2025,Mak2014}. While, polar materials, such as ferroelectrics, have proved to be fertile playgrounds to explore SHG and nonlinear optical effect in general\cite{Denev2011,Nordlander2018,Ok2006,DeLuca2023,Tan2016,Deluca2017}.
Studies on ferroelectric materials have primarily focused on the anisotropy of the even-order harmonic spectrum and its possible connection to the ferroelectric polar axis direction\cite{Gholam2018,Wang2017}.

NbOI$_2$ uniquely belongs to both material classes by exhibiting in-plane ferroelectricity within a van der Waals layered crystal structure. Its ferroelectricity remains stable in thin flakes at room temperature with a remarkably high Curie temperature reaching 189 $^\circ \text{C}$ \cite{Wu2022}. Moreover, the ferroelectric state in NbOI$_2$ has been found to be highly tunable under high pressure\cite{Ye2023}. By virtue of these properties, NbOI$_2$ shows great potential in key applications such as in piezoelectric devices and as conducting domain walls\cite{Cui2024,Hossain2025}.
\begin{figure*}[ht]
    \includegraphics[width=1.0\linewidth]{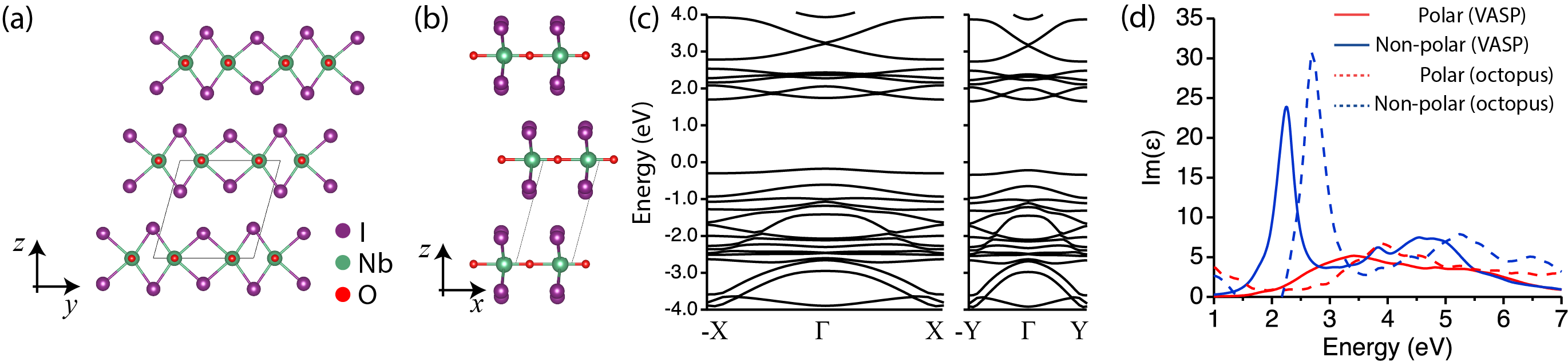}
    \caption{ 
    (a) The crystal structure of NbOI$_2$ along the polar axis (x-axis). (b) The crystal structure of NbOI$_2$ along the non-polar axis (y-axis). The alternating Nb-O bond lengths yield ferroelectricity along the polar axis (x-axis). (c) 
    The DFT band structure with Hubbard $U$ parameters of 3 eV for Nb-d orbitals and 5 eV for I-p orbitals. (d)
    Imaginary part of dielectric function of NbOI$_2$ along the polar and non-polar axes}
    \label{fig:structure-band}
\end{figure*}

The optical nonlinear-response of NbOI$_2$ has also been actively investigated, given the second harmonic, and other even-order harmonics, are directly related to the ferroelectric phase. Previous studies on NbOI$_2$ have mainly focused on examining its giant SHG response and its wavelength and pulse polarization dependence within the perturbative regime\cite{Abdelwahab2022,Ye2023}. Further works emphasize the interplay between SHG and an excitonic resonance at $\lambda_{SHG}\approx 500  \text{ nm}$ \cite{wang2024, Wu2026}. However, the even-order responses of NbOI$_2$ beyond SHG, along with their characteristics in different wavelength regimes remain less clearly understood. 

In this article, we investigate the non-perturbative HHG spectra of NbOI$_2$ in the near-infrared (NIR) and visible regimes using real-time time-dependent density functional theory simulations. We find that higher even-order harmonics exhibit laser polarization dependent anisotropies which are not simply described by the Ferroelectric axis alone. 
Specifically, as the wavelength of incident light is tuned from 2150 nm to 2450 nm,
the second-harmonic remains strongest for excitation along the ferroelectric polar axis, whereas the polarization of the fourth-harmonic changes preferred axis. We propose distinct mechanisms that describe this change in anisotropy and suggest that the angle-dependent response of even-order harmonics is tunable. Our work provides deeper insight into the interplay between ferroelectricity and nonlinear optics in the non-perturbative regime.

\section{Methods}
We used previously reported crystal structure of NbOI$_2$ from single-crystal X-ray diffraction measurements.\cite{Abdelwahab2022} 
Linear-response calculations were carried out using the pseudopotential projector-augmented wave method\cite{Kresse1999} implemented in the Vienna Ab initio Simulation Package (VASP)\cite{Kresse1996,Kresse1996a,Kresse1993}. An energy cutoff of 400 eV was used for the plane-wave basis set. Exchange-correlation effects were treated using the Perdew-Burke-Ernzerhof (PBE) generalized gradient approximation density functional\cite{Perdew1996}. A 16$\times$16$\times$8  $k$-point mesh was used to sample the Brillouin zone. 
Spin-orbit coupling effects were included self-consistently. The DFT+$U$ calculations utilized an effective Hubbard $U$
of 3 eV on the Nb-$4d$ and 5 eV for I-$5p$ states \cite{Dudarev1998}. The Hubbard $U$ values were chosen to reproduce the experimentally observed fundamental band gap and dielectric function. The dielectric function was calculated within the independent-particle approximation \cite{Gajdos2006}, i.e., excitonic effects were not considered.

The transition dipole matrix $|d_x(\mathbf{k})|$ were obtained from WAVEDER file in VASP calculations:

\begin{equation}
|d_\mu (k)|=|\left<u_{n'}(\text{k})|u_{n}(\text{k}+\Delta k)\right>|
\label{eq:dipoleVASP}
\end{equation}
where $u_{n'}(\text{k})$ is the cell-periodic part of pseudo-wavefunctions.

Density functional theory and RT-TDDFT calculations were performed using norm-conserving pseudopotentials\cite{Vansetten2018} as implemented in the Octopus code \cite{Dejean2020}. A 16$\times$16$\times$8 $k$-point mesh was used to sample the Brillouin zone with a real-space grid spacing of 0.17 $\AA$.  Exchange-correlation effects were treated using the PBE generalized gradient approximation density functional\cite{Perdew1996}. Spin-orbit coupling effects and a Hubbard $U$ (same values are in VASP) were included self-consistently\cite{Hamann2013}. A laser pulse duration and intensity of 40 fs and $1.77\times10^{12}$ W/cm$^2$ with a $\cos^2$ envelope were used, respectively. 

\begin{figure*}[ht]
    \includegraphics[width=0.7\linewidth]{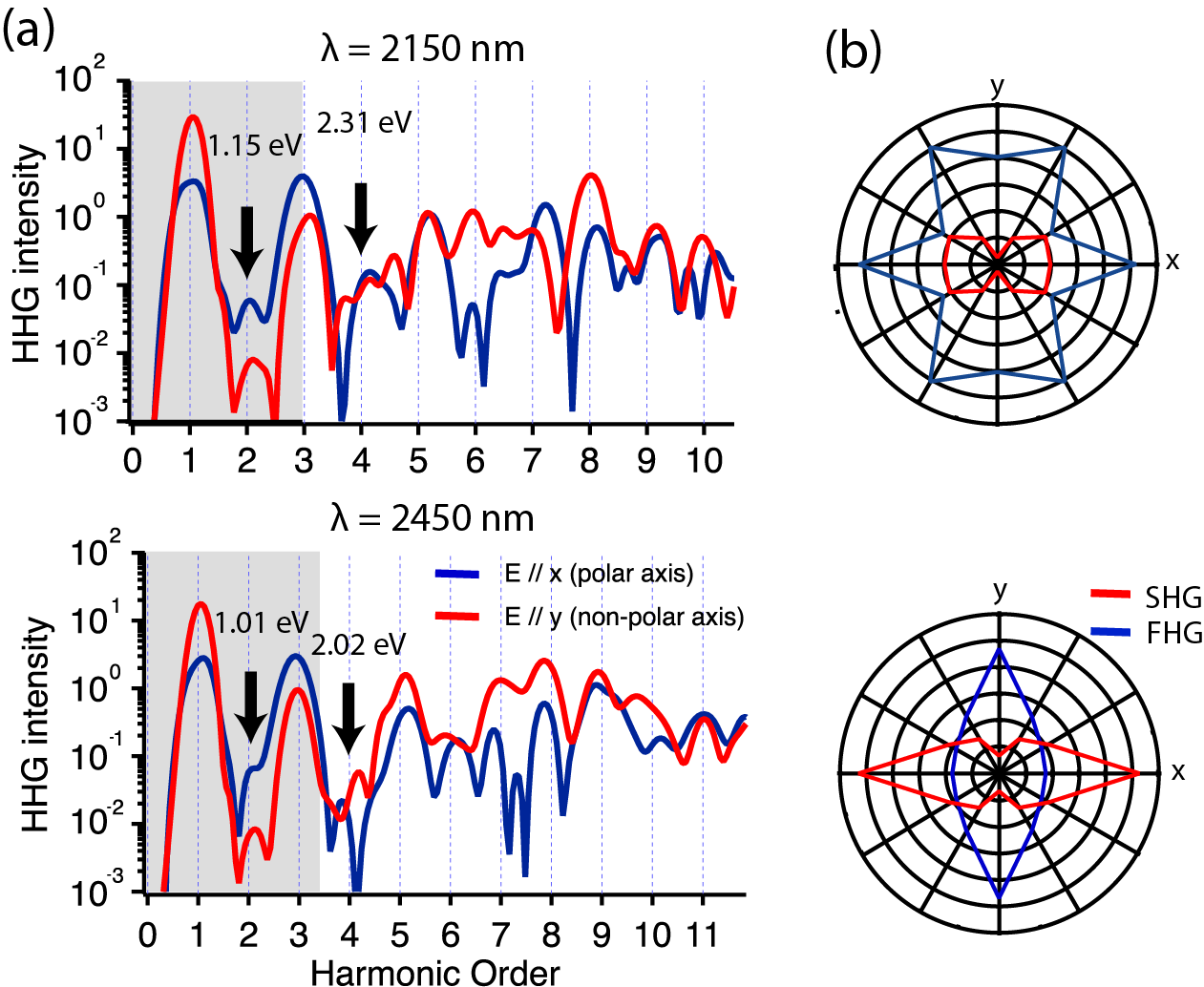}
    \caption{(a) The in-plane HHG spectra of NbOI$_2$ from RT-TDDFT calculations for driving wavelengths of 2150 nm (top) and 2450 nm (bottom). (b) Polar plots of the total in-plane intensity of SHG and FHG as a function of the polarization angle of the laser pulse calculated with 30 $^\circ$ intervals.}
    \label{fig:rttddft-spectra}
\end{figure*}

The HHG spectra were computed from the RT-TDDFT calculations by discrete Fourier transform of the time derivative of total current in the unit cell $\vec{j(t)}$ as given by:
\begin{align}
&\text{HHG}(\omega) = \left|\int_0^T dt~e^{-i\omega t}\frac{\partial \vec{j}(t)}{\partial t}W(t)\right|^2.
\label{eq:HHG}
\end{align}
A Hanning window $W(t)$ was introduced to reduce spectral leakage.

\section{Results and Discussion}
\subsection{Crystal Structure, Electronic Dispersion, and Linear-Response of NbOI$_2$}

\begin{figure*}[ht]
    \centering
    \includegraphics[width=0.8\linewidth]{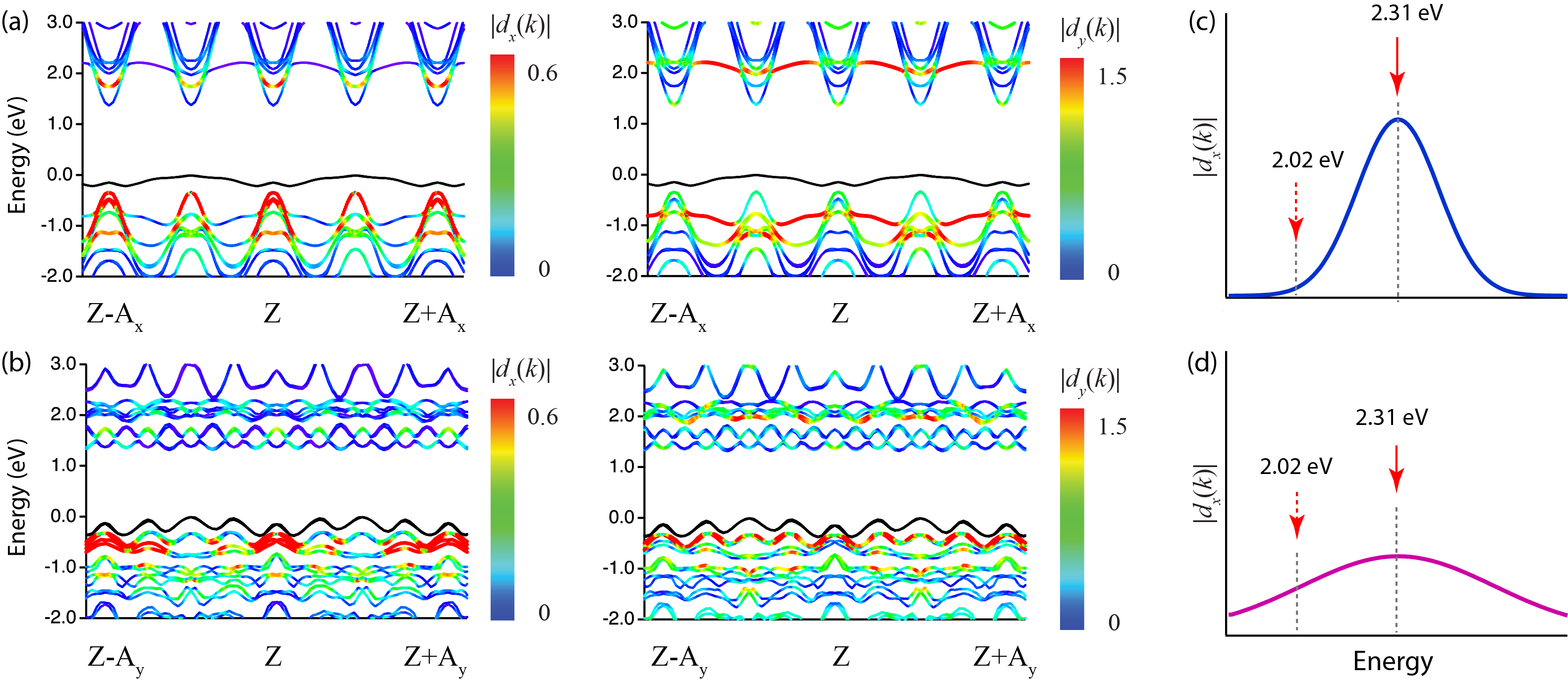}
    \caption{(a) Transition dipole matrix elements, $|d_x(\mathbf{k})|$ (left) and $|d_y(\mathbf{k})|$ (right) from the highest valence band, along the k-path parallel to the polar axis. (b) Transition dipole matrix elements along the k-path parallel to the non-polar axis. Regions with large transition dipole matrix elements are marked by thick red lines. $A_{x}$ and $A_{y}$ represent the vector potential of the 2150 nm laser pulse. (c,d) Schematic figure of the transition dipole distribution for a laser pulse parallel to the polar axis (c) and the non-polar axis (d). The 2.31 and 2.02 eV are the FHG frequencies for wavelengths of 2150 and 2450 nm, respectively}
    \label{fig:transition-dipole}
\end{figure*}

NbOI$_{2}$ is a layered material, where each layer is composed of a Nb metal site coordinated by four planar iodine atoms and two apical oxygen atoms in an octahedral arrangement. Adjacent octahedra share I corners, forming \(\text{Nb–I–Nb}\) chains along the $y$-axis,

whereas Nb-O-Nb bonds along  the $x$-axis bind the \(\text{Nb–I–Nb}\) chains together to form a 2D layer [Fig.\ref{fig:structure-band}(a,b)]. The alternating \(\text{Nb–O–Nb}\) bond length yields the ferroelectricity along $x$-axis. Therefore, we define $x$-axis as the polar axis, and $y$-axis as the non-polar axis.
The 2D atomically-thin layers stack in an alternating fashion along $z$-axis,
where weak van der Waals interactions connect the sheets parallel to the (001) plane, forming a monoclinic crystal structure.

Fig.~\ref{fig:structure-band}(c) presents the electronic band structure of NbOI$_{2}$ along the $x$- and $y$-axis in the Brillouin zone. The band structure exhibits a direct band gap of 1.75 eV consistent with previous DFT calculations using the hybrid functional HSE06 and absorption spectra measurements\cite{Mortazavi2022,Jia2019}. Furthermore, the valence-band ordering, with the flat valence band lying highest in energy among the valence bands, is consistent with angle-resolved photoemission spectroscopy measurements\cite{Abdelwahab2022}. As expected, the band dispersion along the polar axis is symmetric with respect to momentum inversion since they respect both time-reversal symmetry and twofold rotational symmetry about the polar ($x$) axis. Additionally, the band structure along the non-polar axis is nearly symmetric, suggesting intraband contributions to even-order harmonics will nearly vanish, allowing interband components to dominate.

Fig.~\ref{fig:structure-band}(d) shows the imaginary part of the linear-response dielectric tensor with components along the polar (blue) and non-polar (red) crystal axes. Although our dielectric function calculations do not include excitonic effects, the imaginary part of the dielectric function shows overall agreement with previously reported Mueller-matrix ellipsometry measurements, which exhibit excitonic peaks at 2.34 and 2.64 eV along the non-polar axis\cite{Abdelwahab2022}.
This apparent agreement may arise from the underestimation of the band gap within the DFT framework for NbOI$_2$. This interpretation is supported by the large exciton binding energy of 0.63 eV reported in isostructural NbOCl$_2$, \cite{Ding2024} as well as by the proposed excitonic effects in the SHG response of NbOI$_2$\cite{Abdelwahab2022}.

To characterize the excitation regime under
the applied strong laser field, we evaluated
the Keldysh parameter $\gamma$.
The Keldysh parameter describes whether the excitation process is perturbative, involving multiphoton excitation, or non-perturbative, involving Landau-Zener tunneling.\cite{keldysh1964}
\begin{equation}
\gamma = \frac{\omega}{eE}\sqrt{m_r \Delta},
\label{eq:gamma}
\end{equation}
where $\omega$ is the laser pulse frequency, $E$ is the electric field, $m_r$ is the reduced mass of electron-hole pair, and $\Delta$ is the direct band gap. Here, $\gamma =  0.50 $ along the polar axis and $\gamma = 0.61$ along the non-polar
axis, which suggests that the non-perturbative excitation process dominates over the perturbative process. Therefore, the real-time dynamics of the system play an important role beyond a ground-state description.

\subsection{RT-TDDFT Real-Time Dynamics and Interband Coupling Analysis}
To investigate the anisotropic optical response beyond the linear-response and perturbative regimes, we performed RT-TDDFT simulations at various wavelengths. Fig.\ref{fig:rttddft-spectra} presents the HHG spectra for driving wavelengths of 2150 nm (0.58 eV) and 2450 nm (0.51 eV) along with their polarization as a function of angle. In Fig.\ref{fig:rttddft-spectra}a the HHG spectra show distinct features of the second and fourth-harmonics: the second-harmonic remains strong when the laser pulse is aligned along the polar axis throughout the wavelength range, whereas the driving polarization direction that maximizes the fourth-harmonic intensity changes as the wavelength is tuned. We also calculated SHG and FHG intensities with respect to various polarization angles of the laser pulse, and the results are shown as the polar plot in Fig.\ref{fig:rttddft-spectra}b. The polar plot shows that the angle dependence in FHG changes according to wavelength, but the angle dependence in SHG is less sensitive to wavelength. Fig.S2 shows Cartesian-component-resolved spectra for a range of driving wavelengths, which suggest that the current along the polar axis dominantly contribute to the SHG and FHG intensities and their angle dependence. 

Fig.~\ref{fig:transition-dipole} presents the norm of the transition dipole moments $|d_i(\mathbf{k})|$ 
overlaid on the electronic band structure along a k-path (a) parallel and (b) perpendicular to the polar axis, with a schematic (c)-(d) of distribution of transition dipole moments in energy regime. 
Interband HHG intensities can clarify the origin of the harmonic anisotropy since they are closely related to interband coupling among valence and conduction electronic states\cite{Yue2022}. Along the polar axis, the transition dipole $|d_x(k)|$ from the highest valence band is concentrated within a narrow energy window [Fig~\ref{fig:transition-dipole}(a) and (c)]. The corresponding energy regime with concentrated $|d_x(\mathbf{k})|$ is about 2.0 - 2.4 eV [Fig. S4]. In contrast, along the non-polar axis [Fig.~\ref{fig:transition-dipole}(b) and (d)], the transition dipole is distributed over a broader energy range. Such anisotropy in transition dipole supports the anisotropy reversal of FHG. For a laser pulse along the polar axis, the FHG intensity is drastically suppressed when the FHG photon energy moves outside the narrow energy range as the driving wavelength increases. In contrasts, for a pulse along the non-polar axis, the excitations are not suppressed since the distribution of transition dipole matrix elements is broader in energy. 

 Note that the transition dipole matrix elements between valence and conduction bands (Fig.~\ref{fig:transition-dipole}(a)) are dominated by the non-polar axis component, $|d_y(\mathbf{k})|$, rather than the polar axis component, $|d_x(\mathbf{k})|$ along the k-path parallel to the polar axis. This is consistent with imaginary dielectric function in linear response. However, even-order harmonics are allowed by symmetry breaking along polar axis, and dominated by the polar axis components.
 
 Even though RT-TDDFT cannot fully describe the excitonic bound states, the main features of anisotropy can be explained with appropriate energy levels of excited states. It also suggests that higher-order nonlinear processes probe different regions of the band structure and different interband couplings than SHG.

\subsection{Tunability of SHG and FHG}
\begin{figure}[t]
    \centering
    \includegraphics[width=1\linewidth]{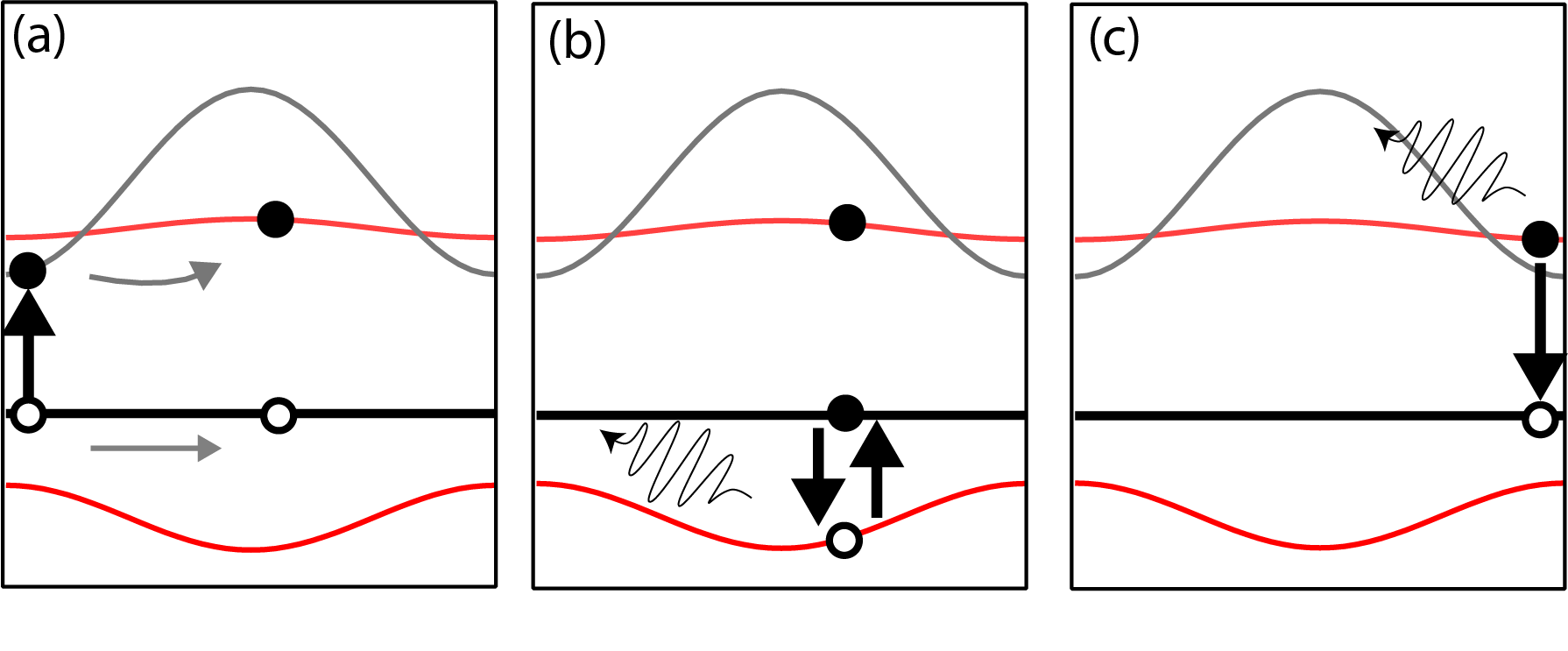}
    \caption{Schematic diagram  of harmonic generation processes. (a) Tunneling at the band edge and acceleration of electron-hole pair. (b) Subsequent excitations and recombinations generating second-harmonics. (c) Recombination between valence and conduction bands generating fourth-harmonics.}
    \label{fig:process}
\end{figure}

Since the frequency of SHG lies below the direct band gap of 1.75 eV, it suggests that the SHG cannot be explained by electron-hole pair recombination between valence-conduction band pairs. In light of our results, we suggest an alternative picture. Fig.~\ref{fig:process} illustrates our picture for the SHG and FHG excitation process, where bands with strong interband coupling are highlighted in red. 
Fig.~\ref{fig:process}(a) shows the creation of electron-hole pair by Landau-Zener tunneling, and acceleration driven by the laser pulse.
SHG in the sub-bandgap regime may arise from transitions between valence bands[Fig.~\ref{fig:process}(b)], which is consistent with large transition dipole matrix between valence bands. In contrast, FHG in the above-bandgap regime is driven by recombination between conduction and valence bands [Fig.\ref{fig:process}(c)].

Based on the distinct mechanisms underlying SHG and FHG, we suggest a pathway for tuning even-order HHG. The anisotropy of FHG is primarily governed by the band width of the narrow conduction state, while SHG is expected to be sensitive to valence-band splitting. By selectively engineering conduction and valence bands, SHG and FHG responses can be modulated independently. For example, selective tuning of FHG can be achieved by controlling the fundamental energy gap through variation of the number of layers or dielectric screening from the substrate. \cite{Raja2017,Ding2024}.

The detailed microscopic processes responsible for individual harmonic orders remain to be clarified. The contributions from other channels, such as conduction–conduction band transitions, can be considered. For example, it has been reported that conduction–conduction band transitions contribute to HHG in the sub-bandgap regime of monolayer NbOCl$_2$ based on semiconductor Bloch equation (SBE) analysis\cite{Li2025}. 

The previous nonlinear optical experiments on NbOX$_2$ (X = Cl, Br, I) focused on SHG and their anisotropies. Instead our results suggest that different even-order harmonics in a ferroelectric material carry different microscopic information. Our work on NbOI$_2$ shows that FHG primarily reveals valence-conduction band coupling and band dispersions, whilst the SHG reveals valence-valence or conduction-conduction band couplings. The difference between SHG and FHG is manifested in their distinct anisotropies with respect to laser pulse polarization angle. Hence, this suggests the tunability of nonlinear optical response in NbOI$_2$.

\section{Conclusion}
In conclusion, we theoretically investigated the angle-resolved even-order high-harmonic response of ferroelectric NbOI$_2$ in the non-perturbative regime for various driving wavelengths in the NIR. By combining RT-TDDFT simulations with transition dipole matrix analysis, we found that the SHG intensity is consistently enhanced when the laser polarization is aligned with the ferroelectric axis. However, the FHG exhibits a stronger frequency dependence of its polarization anisotropy. We suggest the distinct microscopic processes underlying the distinct anisotropies SHG and FHG.

Our results show that different even-order harmonics in NbOI$_2$ provide distinct microscopic information, establishing polarization-resolved HHG as a sensitive probe of anisotropic strong-field dynamics in NbOI$_2$. Moreover, our results suggest that the anisotropy of even-order harmonics is tunable with laser pulse frequency, which provides a pathway for optical manipulation of angle-dependent  even-order harmonics with potential application for optical switches.

\begin{acknowledgments}
This work was carried out under the auspices of the US Department of Energy (DOE) National Nuclear Security Administration under Contract No. 89233218CNA000001. It was supported by the LANL LDRD Program, and in part by the Center for Integrated Nanotechnologies, a DOE BES user facility, in partnership with the LANL Institutional Computing Program for computational resources.
\end{acknowledgments}

\bibliographystyle{apsrev4-2}
\bibliography{references}

\end{document}